\documentclass[10pt]{iopart}
\usepackage{graphicx}
\usepackage{latexsym}
\usepackage{ulem}
\usepackage{dcolumn}
\usepackage{bm}
\usepackage{color}

\newcommand{\be}{\begin{equation}}
\newcommand{\ee}{\end{equation}}
\newcommand{\bea}{\begin{eqnarray}}
\newcommand{\eea}{\end{eqnarray}}
\newcommand{\bse}{\begin{subequations}}
\newcommand{\ese}{\end{subequations}}
\newcommand{\p}{\partial}
\newcommand{\ds}{\displaystyle}
\renewcommand{\comment}[1]{}
\newcommand{\erf}{\mbox{erf}}

\begin{document}

\paper[]{Relaxation and coarsening of weakly-interacting breathers in a simplified DNLS chain}
\date{\today}

\author{Stefano Iubini$^{1,2}$, Antonio Politi$^{3}$, Paolo Politi$^{4,2}$}

\address{$^{1}$ Dipartimento di Fisica e Astronomia, Universit\`a di Firenze, via G. Sansone 1 I-50019, Sesto Fiorentino, Italy}
\address{$^{2}$ Istituto Nazionale di Fisica Nucleare, Sezione di Firenze, via G. Sansone 1 I-50019, Sesto Fiorentino, Italy}
\address{$^{3}$ Institute for Complex Systems and Mathematical Biology \& SUPA
           University of Aberdeen, Aberdeen AB24 3UE, United Kingdom}
\address{$^{4}$ Istituto dei Sistemi Complessi, Consiglio Nazionale
delle Ricerche, via Madonna del Piano 10, I-50019 Sesto Fiorentino, Italy
}

\ead{stefano.iubini@unifi.it}

\begin{abstract}
The Discrete NonLinear Schr\"odinger (DNLS) equation displays a parameter region characterized by
the presence of localized excitations (breathers). While their formation is well understood and 
it is expected that the asymptotic configuration comprises a single breather on top of a background,
it is not clear why the dynamics of a multi-breather configuration is essentially frozen.
In order to investigate this question, we introduce simple stochastic models, 
characterized by suitable conservation laws.
We focus on the role of the coupling strength between localized excitations and background.
In the DNLS model, higher breathers interact more weakly, as a result of their faster rotation.
In our stochastic models, the strength of the coupling is controlled directly by an
amplitude-dependent parameter.
In the case of a power-law decrease, the associated coarsening process undergoes a slowing down
if the decay rate is larger than a critical value. In the case of an exponential decrease,
a freezing effect is observed that is reminiscent of the scenario observed in the DNLS.
This last regime arises spontaneously when direct energy diffusion between breathers and background 
is blocked below a certain threshold.

%Finally a more general stochastic model is introduced, where an exponentially decreasing interaction
%emerges spontaneously.
\end{abstract}

\submitto{Journal of Statistical Mechanics: theory and experiment}

\section{Introduction}

The one-dimensional Discrete NonLinear Schr\"odinger (DNLS) equation~\cite{Eilbeck1985,Kevrekidis}
\be
\label{eq.DNLS}
i \dot {z}_n = -2 |z_n|^2z_n - \gamma( z_{n+1}+z_{n-1}) \, 
\ee
with $1\le n\le N$ and suitable boundary conditions 
describes the dynamics of a chain of interacting, anharmonic oscillators with complex amplitudes $z_n$.
This equation was firstly derived in the 1950s by Holstein~\cite{Holstein} within a tight-binding approximation
for the motion of polarons in molecular crystals, but it can more generally be derived by
a nonlinear Schr\"odinger equation with a simple cubic nonlinear term, once we discretize the
Laplacian as a finite difference approximation.
Since then, because of its generality, it has become a prototypical model of wave 
propagation in nonlinear lattices~\cite{Hennig99} and 
provides a reasonably accurate description of several physical setups, ranging
from trapped cold gases~\cite{trombettoni2001,Hennig2016dynamical,Franzosi2011} to coupled optical 
waveguides~\cite{jensen1982,christodoulides1988}, from magnetic systems~\cite{rumpf2001,borlenghi14,borlenghi15}
to energy transport in biomolecules~\cite{Scott2003}.

Upon identifying the set of canonical variables $z_n$ and $iz_n^*$, Eq.~(\ref{eq.DNLS})  follows from 
the Hamilton equations $\dot z_n = -\partial H/\partial (iz_n^*)$ associated to the Hamiltonian
\be
\label{eq:general1}
 H = \sum_{n=1}^N \Big[ |z_n|^4 + \gamma( z_n^*z_{n+1}^{}+z_n^{}z_{n+1}^* ) \Big] , 
\ee
which is a conserved quantity along with the mass
\be
\label{eq:A}
A = \sum_n |z_n|^2 \; .
\ee

The thermodynamical behavior of the DNLS equation depends on two main parameters: the energy density $u=H/N$ and the mass density 
$a=A/N$, which can be mapped unambiguously to a couple of values of  temperature $T$ and chemical potential $\mu$.
In the limit of a vanishing temperature, the lowest energy state can be determined by searching for
a uniform
%translationally invariant configuration, i.e.  a 
state with constant amplitude $|z_n|$ 
and a constant phase shift between neighbouring sites, $z_{n+1}=z_n \exp(i\phi_0)$.
Posing $z_n = |z|\mathrm{e}^{i(n\phi_0+\psi)}$, where $\psi$ is an irrelevant $n-$independent quantity, 
it is found that $u=|z|^4 + 2\gamma |z|^2\cos\phi_0$ and $a= |z|^2$. 
The minimization gives $\phi_0=\pi$, so that the ground states lie along the line
$\mathcal{C}_0$, $u=a^2 -2\gamma a$. States below $\mathcal{C}_0$ are not physically accessible.
Moreover, by assuming $\psi=\omega_0 t$, it is found from Eq.~(\ref{eq.DNLS}) that a ground state
rotates with a frequency $\omega_0=2(a-\gamma)$, which depends on its mass density.

In the limit of a diverging temperature, the coupling term on average vanishes as the phases are
randomly distributed. Therefore, the mean frequency, $\omega= 2|z_n|^2$, is proportional 
to the local mass, which is distributed according to a Poisson distribution~\cite{Rasmussen2000}.
In fact, in the limit $\beta\to 0$ ($T\to\infty$) the statistical weight of a configuration,
given by the grand canonical partition function, is proportional 
to $\exp(\eta A)$, where $\eta=\beta\mu$ is the (negative) finite limit of the product of the vanishing
inverse temperature $\beta$ by the diverging chemical potential $\mu$.
In conclusion, infinite-temperature states are characterized by an exponential distribution
$P(|z_n|^2)=a^{-1}e^{-|z_n|^2/a}$, with $\langle |z_n|^2\rangle=a$, so that they lie along the line
$\mathcal{C}_\infty$,  $u=\langle |z_n|^4\rangle = 2a^2$.

Equilibrium states in between $\mathcal{C}_0$ and $\mathcal{C}_\infty$ are {\it standard} positive-temperature
thermodynamic states, while  the points above $\mathcal{C}_\infty$ belong to the so-called negative temperature 
region~\cite{Rasmussen2000}. Although  a grand canonical distribution is ill-defined for the
negative-temperature DNLS~\cite{Rasmussen2000,johansson2004}, negative temperature states can be consistently introduced within the
microcanonical ensemble~\cite{johansson2004}, where $H$ and $A$  are fixed  quantities.  
%are characterized by negative absolute temperatures,
%because the injection of energy reduces the entropy,
%as expected for any energy spectrum which is bounded from above, once all levels are equally 
%populated.
The main feature of the negative temperature regime in the DNLS equation is the spontaneous appearance of
localized excitations, called breathers~\cite{flach2008}, which collect the excess energy density
$(u-2a^2)$. Since the local  frequency of the phase of each oscillator is proportional
to its mass, breathers are expected to rotate at a much larger frequency than other sites, usually named background.
Direct microcanonical  simulations of the DNLS chain show that, after some transient, the breather dynamics is essentially
frozen for very long times~\cite{iubini2013},
with a finite density of breathers which is constant in time and a negative microcanonical temperature. 
This is in contrast with analytical results based on entropic 
arguments~\cite{Rumpf2004,Rumpf2007,Rumpf2008,Rumpf2009},
whose conclusions lead to expect that the density of breathers
should instead decrease in time and that the expected final (equilibrium) state is
made up of a single breather, which absorbs all the excess energy, surrounded by an infinite-temperature 
background.

\begin{figure}
\begin{center}
\includegraphics*[width=0.8\textwidth]{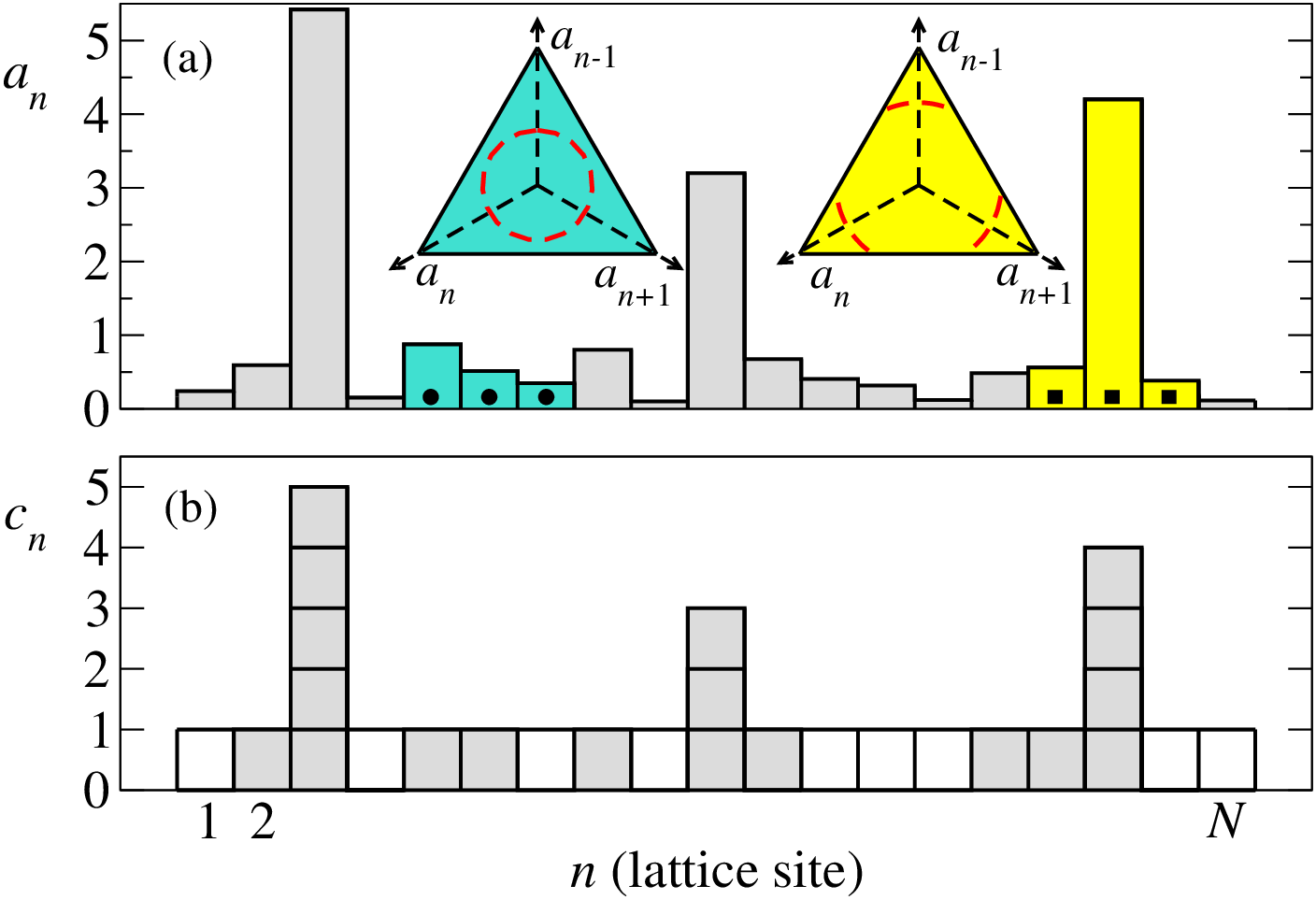}
\end{center}
\caption{
(a): The MMC model. A typical amplitude profile $a_n$ displays a certain number of high-amplitude 
breathers superposed to a background. 
Left and right insets show in dashed red lines the available states respectively for a
background triplet of sites (turquoise bars marked with black dots) and a triplet containing a breather (yellow bars
marked with black squares).
The legal configurations are obtained as intersection between a plane (main triangle) and a sphere (not shown)
that account for local conservation of mass and energy, respectively.
Due to the further constraint on the positivity of the amplitudes $a_n$, the allowed states may
lie on three disconnected arcs, see the right inset.
This happens when one of the three amplitudes is significantly larger than the others.
(b): The PEP model. The amplitude profile in panel (a) is reproduced in terms of the integer amplitude $c_n$.
White and grey squares highlight respectively empty ($c_n=0$) and occupied ($c_n>0$) sites.
}
\label{f:MMC_PEP}
\end{figure}

A full understanding of this discrepancy in the deterministic DNLS model has not yet
been reached, due to the difficulty of providing a consistent description of the interaction between 
a breather and the nearby fluctuating background over very long (i.e. thermodynamical) timescales.
As a result, the problem of determining the macroscopic relaxation law that allows to reach the final single-breather state is still open.
In order to gain insight into this problem, in this paper we study some simplified stochastic versions of 
the DNLS equation, as their dynamics can be much 
better controlled and characterized. These models have been selected so as to exhibit conservation laws,
since we are convinced that these are key properties of the general scenario.
A base version of the two main models was introduced in a previous publication~\cite{IPP14}; it is here
illustrated in Fig.~\ref{f:MMC_PEP}, where we also stress the mutual relationships.
The first model, panel (a), is obtained by neglecting the coupling term, i.e. by setting $\gamma=0$
in the DNLS Hamiltonian. Since the critical line $\mathcal{C}_\infty$ does not depend on the value
of $\gamma$, we expect it to still hold in the stochastic model.
In the limit of vanishing $\gamma$, the  local phases of the DNLS oscillators do not play any role.
Accordingly, it is convenient to introduce the local variable $a_n=|z_n|^2$ denoting the mass (or ``height") of the site $n$
and the two conserved quantities~(\ref{eq:general1}) and ~(\ref{eq:A})  write as $H=\sum_n a^2_n$ and $A=\sum_n a_n$.
The stochastic model dynamics is finally defined by introducing a local Microcanonical Monte Carlo (MMC) move.
Given a randomly chosen triplet of consecutive sites, $(n-1,n,n+1)$, the local variables 
$a_{n-1},a_n,a_{n+1}$ are randomly updated under the only constraint of preserving the total energy and mass 
of the triplet. In the following we will refer to this model as the MMC model.

The second model, panel (b), can be considered as a further simplification of the DNLS dynamics and 
is better explained by distinguishing between breather and non-breather sites
(also called background sites). The height is now an integer variable that we call $c_n$ in order to mark the difference with
respect to the previous setup. 
In the background $c_n$ can only take the values $0$ (empty site) or $1$ (occupied site), while breathers are characterized
by $c_n>1$. The model is called partial exclusion process (PEP), 
because the space is divided into disjoint channels (the regions among breathers) where
particles diffuse according to the exclusion constraint. Breather
sites instead can freely emit and absorb particles towards/from the neighbouring background sites,
until their height reduces to one, in which case they are absorbed by the background.

In both models, above the infinite temperature line, a slow dynamics manifests itself as a
sort of coarsening of breather states, which has, however, no equivalent in the DNLS, where their dynamics essentially freezes.
So, the question arises: what is the relevant ingredient which is missing in the stochastic models?
An important, conceptual difference between the DNLS equation and the MMC/PEP models is the
absence of a phase dynamics in the latter ones. In the original context, the local variable is indeed
$z_n := \sqrt{a_n} \mathrm{e}^{i\phi_n}$, while only $a_n$ is present in the MMC.
This is contrary to the approximation often invoked in the context of dissipative oscillators,
where the amplitude dynamics is neglected, the phases being much more sensitive to the coupling strength~\cite{Kuramoto}. 
The approach is, however, justifiable in our
context when $\gamma$ is small: in this limit, neither of the two conserved quantities depends on the phases, 
so that the physics is contained in the distribution of the amplitudes.
Phases enter only in the coupling mechanism, which, in the MMC, has been designed just to 
capture entropic effects in the simplest possible way. 
As a result, one can claim that the disagreement between the coarsening dynamics (exhibited by MMC and PEP)
and the evolution of breathers in the DNLS can be traced back to the way breather-background interaction
is accounted for. 
More specifically, in the original DNLS, breathers of increasing amplitude rotate faster and faster
(the frequency $\omega$ of a massive breather is equal to $2 |z_n|^2= 2 a_n$).
Therefore, the average coupling energy
becomes increasingly weak upon increasing $a_n$. An explicit perturbative analysis of the DNLS is a subtle object 
that is currently under investigation. Here, we focus on simple stochastic models,
where we have a full control of the evolution rule.

The weakening of the interaction induced by an increasingly fast rotation is here simulated
by postulating that the coupling strength depends on the breather amplitude, which, from now on, is
denoted with $h$. More precisely, we introduce the probability $\alpha \in (0,1]$ 
for a move involving the breather to actually occur. After investigating the case of constant $\alpha$ 
(in order to test the correct scaling) we consider the more interesting case $\alpha = h^{-\beta}$.
One of the major results of this paper concerns the coarsening exponent $1/\zeta$, defined from the time
dependence of the average distance $L$ between neighbouring breathers: $L(t) \sim t^{1/\zeta}$. We find
\begin{equation}
\zeta  =  \left\{
\begin{array}{cc}
\ds 3  & \beta\le 1 \\
\ds 2+\beta  & \beta>1
\end{array} \right. .
\end{equation}

Altogether, in section 2, we analyse the relaxation of a single breather for fixed small $\alpha$ in the PEP
model, finding that the process is initially ballistic, and becomes diffusive at later times.
A similar scenario is then found when $\alpha$ is assumed to depend on the breather height.
In the following section 3, we focus on the interactions between neighbouring
breathers, showing that the process is eventually diffusive. We also connect the value of the diffusion
coefficient with the coarsening exponent. In section 4, we analyse more natural coupling schemes
in the MMC, including one which leads to a logarithmically slow process, quite close 
to the scenario actually observed in the DNLS. Finally, in section 5, some conclusions are drawn
and the open problems briefly summarized.

\section{Relaxation of a single weakly-interacting breather}

In this section we study the PEP model defined on a lattice of $N$ sites with periodic boundary conditions,
see Fig.~1(b).
The variable $c_n$ identifies the number of particles in the site $n$, which can be of
background or breather type. In the former case, 
the particles diffuse as in a standard exclusion process, i.e. $c_n$ can be at most equal to 1.
The breathers are ``reservoirs" ($c_n > 1$) which exchange particles with neighbouring
(background) sites. If the breather content reduces to just a single particle (i.e., $c_n=1$)
it becomes a background site for ever.
The evolution rule is simple: we randomly choose an ordered pair of neighbouring sites $(i,j)$, $j=i\pm 1$, 
and make the
move $(c_i,c_j) \to (c_i -1,c_j+1)$ if and only if $c_i > 0$ and $c_j \ne 1$. This means that we move a particle if
it exists and enters either an empty background ($c_j=0$), or a breather site ($c_j >1$).
If the move involves a breather (i.e., either $c_i > 1$ or $c_j >1$), it is accepted with probability $\alpha$; 
if it does not, it is always accepted. 

At variance with both MMC and the original DNLS, in this model there is only one conserved quantity (the number
of particles), accompanied by an additional constraint in the background, due to the exclusion rule.
There is another difference between PEP and MMC/DNLS models:
in the former one breathers do not arise spontaneously. In fact, according to above rules a breather can be
destroyed but it cannot be created. 
This notwithstanding, the PEP model exhibits a slow dynamics (i.e. the coarsening of breathers)
very similar to that of the MMC model, if the overall density of particles $c$ is larger than 1/2~\cite{IPP14}.
In this respect,
it is instructive to compare the critical density $c=1/2$ of the PEP model with the infinite-temperature
line $u=2a^2$ of the MMC. For $T=\infty$, the MMC is characterized by a Poissonian distribution of masses, 
i.e. $\langle a^2\rangle -\langle a\rangle^2 = \langle a\rangle^2$.
In the PEP model, where $a$ is a binary variable with average $c$, the above condition
writes $c-c^2 = c^2$, whence $c=1/2$.

\begin{figure}[ht]
\begin{center}
\includegraphics*[width=0.8\textwidth]{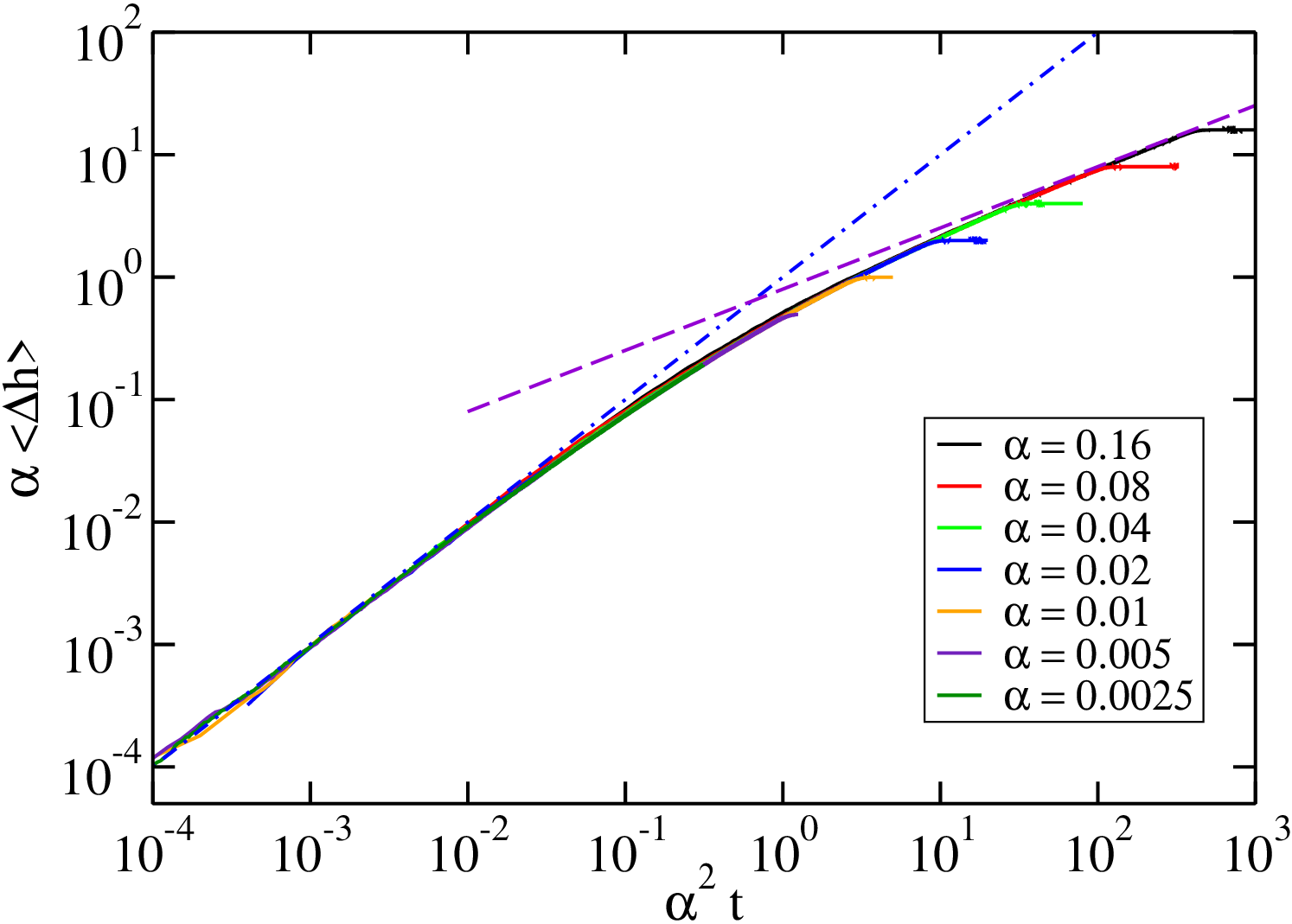}
\end{center}
\caption{Evolution of the average height variation $\langle h(0)-h(t) \rangle$ of the breather
during the relaxation process in rescaled units. Averages have been performed over 2000 different
initial conditions.
The dotted-dashed blue curve represents the analytical prediction for the ballistic growth,
Eq.~(\ref{PEP_small_t}), and the dashed purple curve is the analytic prediction for the diffusive
growth, see Eq.~(\ref{PEP_large_t}).
}
\label{f:relax_PEP}
\end{figure}

In Fig.~\ref{f:relax_PEP} we report the average evolution of a single breather of initial height 
$h(0) < N/2$ for different values of the coupling $\alpha$.
The breather is initially sitting on an empty background with periodic boundary conditions 
and the condition on $h(0)$ implies that the breather is 
eventually absorbed by the background\footnote{Indeed, the condition $h(0) < N/2$ corresponds to a global density
$c<1/2$, which can not sustain localized states when thermodynamical  equilibrium is reached.}.
%because the infinite-temperature regime corresponds to an occupation
%density equal to $1/2$.
In detail, we have chosen $h(0)=100$, much smaller than $N=10^4$,  to reduce boundary effects during 
the breather relaxation. The results show the existence of two regimes separated by a crossover time 
$t_c \sim 1/\alpha^2$. 
For short times the average height of the breather decreases
ballistically, $\langle\Delta h (t)\rangle \equiv \langle h(0) - h(t)\rangle \simeq \alpha t$, while for large times it decreases diffusionally,
$\langle\Delta h(t)\rangle \simeq \sqrt{t}$. 
At even longer times $\langle \Delta h(t) \rangle$ saturates because of the finite height of the breather, which
runs out of particles. 
The different behavior at short/long times can be qualitatively understood as follows.
At early times (especially if $\alpha$ is vanishingly small) released particles 
freely diffuse in a practically empty background with no mutual interaction and
a vanishing probability to be reabsorbed. At long times, emitted particles have a much 
higher probability to return to the breather.

This argument can be made more rigorous under the approximation of continuous time and space variables.
We proceed into two steps, by first deriving a set of mean-field differential equations for
the probability $p_n(t)$ that the site $n$ is occupied at time $t$ (the breather being located in the site
$n=0$),
\bea
\hskip -2.cm \dot p_n \equiv \frac{\Delta p_n}{\Delta t} &=& \frac{1}{2} 
\left [p_{n-1}(1-p_n) + p_{n+1}(1-p_n) - p_n(1-p_{n-1}) 
-p_n(1-p_{n+1})
\right] = \nonumber \\
\hskip -2.cm&=& \frac{1}{2} (p_{n+1} + p_{n-1} - 2p_n) \qquad n > 1
\label{eq:diffdis}
\eea
and
\bea
\hskip -2.cm \dot p_1 \equiv \frac{\Delta p_1}{\Delta t} &=& \frac{1}{2} \left [\alpha (1-p_1) -\alpha p_1 +  
                   p_2(1-p_1) - p_1(1-p_2) 
\right ] \nonumber \\
\hskip -2.cm &=& \frac{1}{2} \left [\alpha (1-2p_1) + (p_2 - p_1)\right] ,
\eea
where $\Delta t =1$ corresponds to the implementation of $N$ random moves ($N$ is the lattice size).
The evolution equation for $p_1$ can be made formally equivalent to the bulk dynamics (\ref{eq:diffdis}) 
upon introducing a $p_0$ such that
\be
 \alpha (1-2p_1) + (p_2-p_1) = (p_0+p_2) -2p_1 \; .
\label{eq:bc}
\ee
As a next step, we introduce the continuous variable $x=n-1$ and the corresponding probability density
$\rho(x,t) = p_n-1/2$, so that the stationary solution is $\rho(x)=0$ . 
Under the approximation of a weak dependence of $\rho$ on the spatial variable $x$ (which becomes increasingly
correct at long times), the bulk dynamics
is described by a standard diffusive equation 
\be
\partial_t \rho(x,t) = D \partial_{xx} \rho(x,t) \; ,
\label{eq:dif1}
\ee
where $D=1/2$. Moreover, one can assume\footnote{Let us remind that $\rho(0) +\frac{1}{2}$ corresponds,
in the discrete lattice, to $p_1$.} 
$p_0 = \rho(0) +1/2 - \rho_x(0)$,  so that
Eq.~(\ref{eq:bc}) transforms into the boundary condition
\be
\label{eq:robin}
\rho_x(0) = 2\alpha \rho(0)  \equiv r \rho(0) .
\ee
Therefore, we recover the well known result that the exclusion process is purely diffusive~\cite{KRB}  and
find that the interaction with the breather can be modelled by a Robin (semi-reflecting) boundary condition
in $x=0^+$, gauged by the variable $r$.
For $\alpha \to\infty$, the Robin condition reduces to a standard absorbing boundary condition, $\rho(0)=0$, 
while for $\alpha \to 0$ , it corresponds to a reflecting boundary, $\rho_x(0) = 0$. 
The case $\alpha=1$, studied in Ref.~\cite{IPP14}, corresponds to an intermediate setup, 
characterized by a finite interaction time-scale, because for $\alpha=1$ attachment occurs on the
time scale of diffusion.

We now want to determine the average height reduction $\langle \Delta h(t)\rangle $ of a breather due to the particles that
have been emitted but {\it not yet} reabsorbed. To this end,  we need to know the probability $F(t)dt$ that a particle 
is being reabsorbed in the time interval $(t,t+dt)$.  This quantity can be evaluated exactly using the continuum model and
assuming that a particle is released in $x=x_0$ at $t=0$ and it is thereby let free to diffuse in $[0,\infty]$.
Note that the value of $x_0$ is not crucial for the continuum model as long as it is chosen to be close to the origin. 
%In the discrete model the emission point is not crucial so long as it is close to the origin.
The analytical expression for $F(t)$ valid for all $t$  is fairly complicated and it is given in~\ref{app.Robin}. Here we
limit to give its expression for short and long times,
\begin{equation}
\label{eq:F(t)}
F(t) \simeq  \left\{
\begin{array}{cc}
\ds 4 \alpha \sqrt{D/(\pi t)} & x_0^2 \ll t \ll \alpha^{-2} \\
\ds 1/(2\alpha\sqrt{\pi Dt^3})  & t\gg \alpha^{-2}
\end{array} \right. .
\end{equation}

The height reduction $\langle \Delta h(t)\rangle$ of the breather is thereby obtained by integrating  over all
particles that have been emitted and not yet absorbed at time $t$. 
However, we must be careful, because the evaluation of the number of emitted
particles (either reabsorbed or not) is not trivial: the emission 
of a particle from the breather is possible only if the neighbouring site, which should
receive it, is empty. Since the initial condition corresponds to an empty background,
at short times the probability $(1-p_1)$ that the neighboring sites are empty is practically equal to 1.
On the other hand, at long times, after many emissions of particles, the region around the breather is
almost at equilibrium, corresponding to $p_1=1/2$.
In general, one can write 
\be
\label{eq:h(t)}
\langle \Delta h(t)\rangle  = \alpha \int_0^t dt' (1-p_1(t')) \int_{t'}^{\infty} dt'' F(t'').
\ee
Using the limiting expressions for $F(t)$ (provided in Eq.~(\ref{eq:F(t)})) and for $p_1(t)$
(discussed here above), we find the following two regimes,
\bea
\label{PEP_small_t}
\langle \Delta h(t)\rangle &\simeq& \alpha t \qquad x_0^2 \ll t \ll \alpha^{-2} \\ 
\langle \Delta h(t)\rangle &\simeq & \sqrt{\frac{2t}{\pi}} \qquad\qquad t \gg \alpha^{-2} \, ,
\label{PEP_large_t}
\eea
where we have used $D=1/2$. 
The details of the derivation of Eqs.~(\ref{PEP_small_t}) and~(\ref{PEP_large_t}) 
are given in~\ref{app.relax}. These limiting behaviors are plotted in
Fig.~\ref{f:relax_PEP} and show a very good agreement with numerical simulations. 

\begin{figure}[ht]
\begin{center}
\includegraphics*[width=0.8\textwidth]{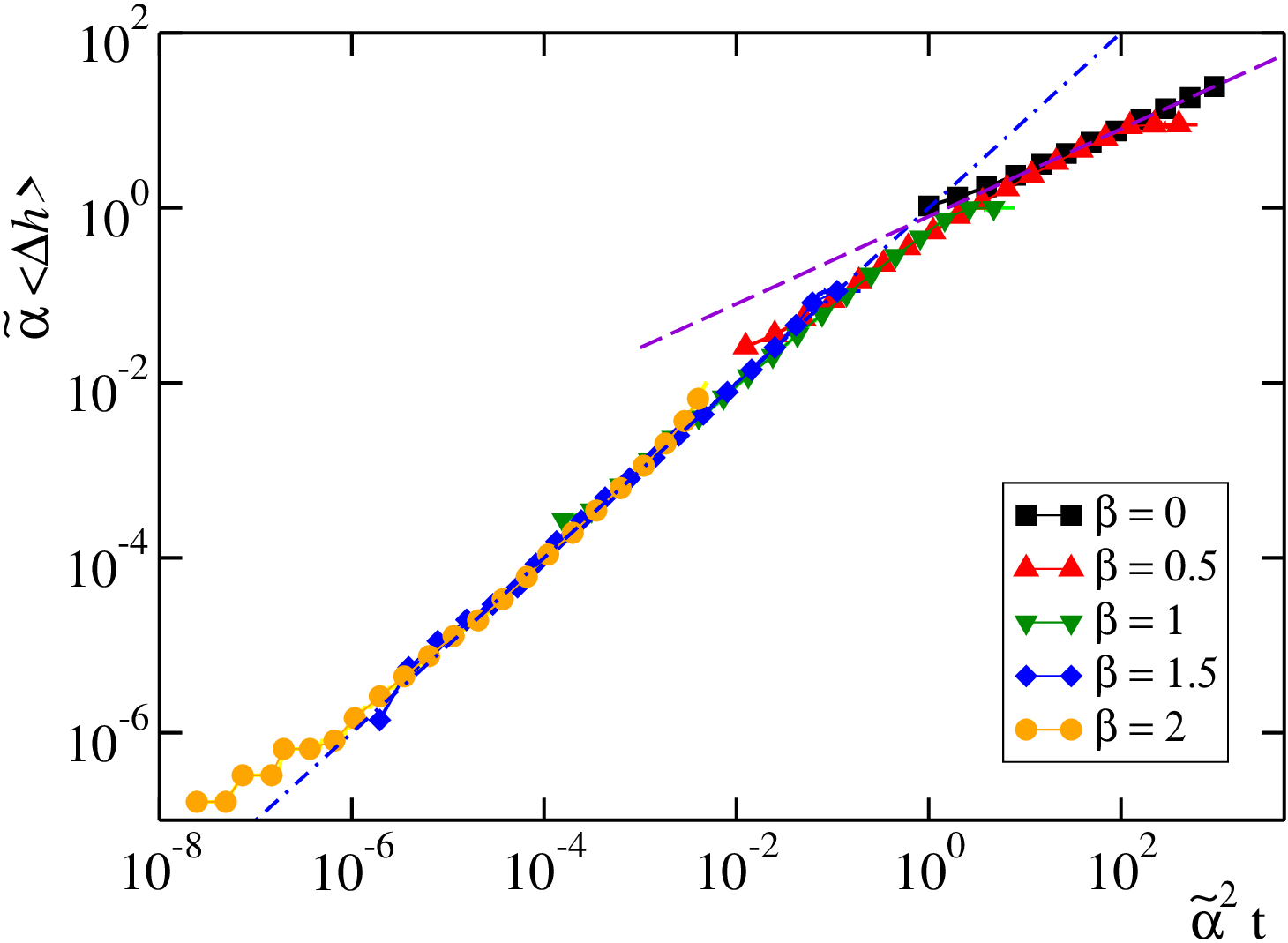}
\end{center}
\caption{Breather relaxation dynamics for different choices of the coupling exponent $\beta$. Simulations refer to 
a setup with $N=10000$ lattice sites and a breather with initial height $h(0)=80$. The PEP channel is initially empty.
Data are averaged over 1000 realizations.
The dotted-dashed and dashed curves represent the analytical predictions as in Fig.~\ref{f:relax_PEP}, with $\alpha$
replaced by $\tilde \alpha$.
}
\label{f:relax_PEP_beta}
\end{figure}

So far we have considered a constant coupling strength $\alpha$. We now turn to a more physical 
case where $\alpha$ varies with the breather height
\[
\alpha(h) = h^{-\beta} \,
\]
and $\beta$ is a real and positive parameter. The dependence of $\alpha$ on $h$ is such that 
the higher a breather is, the lower is its coupling with the background. This also implies that 
the effective coupling is implicitly time-dependent.
The data obtained for different $\beta$-values are reported in Fig.~\ref{f:relax_PEP_beta}, using the same
scaling ansatz as before, with the only difference that now, $\alpha$ is referred to the initial amplitude
(i.e. $\tilde \alpha = \alpha(h(0))$).

We interpret this result as an indication that even in the regime where $\alpha$ decreases with a power of $h$,
the relaxation of the breather displays essentially the same behaviour of the case of $\alpha$ constant. 
The good data collapse obtained in Fig.~\ref{f:relax_PEP_beta} shows that the relevant timescale of the overall
process is the slowest one, namely ${\tilde \alpha}^{-2}$.
\section{Breather interactions}

Here we are mostly interested in determining the exponent $\zeta$ which controls the way the density
$d_B$ of breathers decays in time ($d_B \approx t^{-1/\zeta}$). 
In order to do so, it is first necessary to understand how breathers interact with each other.
After some transient, a set of breathers is present, which sit on a background
characterized by the occupation density $\rho=1/2$.
Let the origin of the time variable be set at such a stage, when the average distance between neighbouring
breathers is $L$, while their average height is $\overline h = kL$, with $k>0$ (this ensures that the surplus of
energy contained in the breathers is an extensive quantity).
In these conditions the background is ``in equilibrium" at $T=\infty$ 
and, on average, the breathers do neither absorb nor release particles.
However, the particles stochastically emitted can occasionally diffuse and be absorbed by a neighbouring 
breather. This mechanism couples breathers, 
which can exchange particles through the background, therefore inducing a diffusion of breathers' height.
Let us focus on a couple of breathers with initial height $h(0)=kL$.  The square displacement $\sigma^2$  
\be
\sigma^2 \equiv \langle (h(t)-h(0))^2\rangle
\ee
is expected to grow as $\sigma^2 = D_B t$, where $D_B(L)$ is the diffusion constant of the exchange process.
By definition, one of the two  breathers is completely absorbed after a time $\tau$, when 
$\sigma^2(\tau)\simeq h^2(0)$. Therefore the absorption time $\tau$ scales with $L$ as $\tau(L) \simeq h^2(0)/D_B(L)$.
In order to determine the coarsening exponent  it is necessary to invert the relation $d_B \approx \tau^{-1/\zeta}$,
once set $d_B=1/L$ and $h^2(0)\simeq L^2$. Accordingly,
\be
L \simeq \left(\frac{L^2}{D_B} \right)^{1/\zeta} \quad.  
\ee
As a result, the coarsening exponent is fully determined by the scaling of $D_B$ with $L$.

%The scaling of the absorption time $\tau$ with $L$ 
%\be
% \tau \approx L^{1/\zeta} = d_B^{-1/\zeta}
%\label{eq:zeta}
%\ee
%gives the exponent $\zeta$ since $d_B\sim L^{-1}$
%In order to determine $\zeta$, it is necessary to first determine the diffusion coefficient $D_B$.
%It basically corresponds to the probability (per unit time) for a breather to pass a particle to a neighbouring breather.
With reference to the two-breather setup, $D_B$ can be expressed as the product of the rate $\gamma$ 
to release a particle tout court  
by the probability $P_c$ that the particle is absorbed by the neighboring breather 
(instead of being absorbed back by the emitter).
An analytical calculation of $P_c$ is reported in~\ref{app.rw} for a simple geometry consisting of only
two breathers placed at the boundaries of a PEP channel with fixed boundary conditions.
For this setup,
\be
\label{eq:zeta}
\gamma=\frac{\alpha}{2}, \qquad
P_c(L,\alpha) = \frac{1}{2(1+L\alpha)} 
\ee
and therefore
\be
D_B(L,\alpha)  = \frac{\alpha}{4(1+L\alpha)}.
\ee
When $\alpha$ scales with the breather height, 
i.e. when $\alpha=\bar h^{-\beta} = (kL)^{-\beta}$, the above equation provides the relevant 
scaling of $D_B$ with the system size $L$. In particular, we find
\be
\label{eq:diff_const}
D_B = \left\{ 
\begin{array}{cc}
\displaystyle \frac{S(k,\beta,L)}{L} & \beta \leq 1 \\
\displaystyle \frac{S(k,\beta,L)}{L^\beta} &  \beta > 1
\end{array}
\right. ,
\ee
where $S(k,\beta,L)$ is a prefactor weakly dependent on $L$ for large $L$. Explicitly,
\be
\label{eq:pref}
S(k,\beta,L)= \left\{ 
\begin{array}{cc}
\displaystyle \frac{k^{-\beta}L^{1-\beta}}{4\left(1+k^{-\beta}L^{1-\beta} \right)} & \beta \leq 1 \\
\displaystyle \frac{k^{-\beta}}{4\left(1+k^{-\beta}L^{1-\beta}  \right)} &  \beta > 1
\end{array}
\right.
\ee
Altogether, from Eq.~(\ref{eq:zeta}) we finally obtain $\zeta=3$ for $\beta\leq 1$ and $\zeta=(2+\beta)$ for
$\beta>1$.

\begin{figure}[htbp]
\begin{center}
\includegraphics*[width=0.8\textwidth,clip]{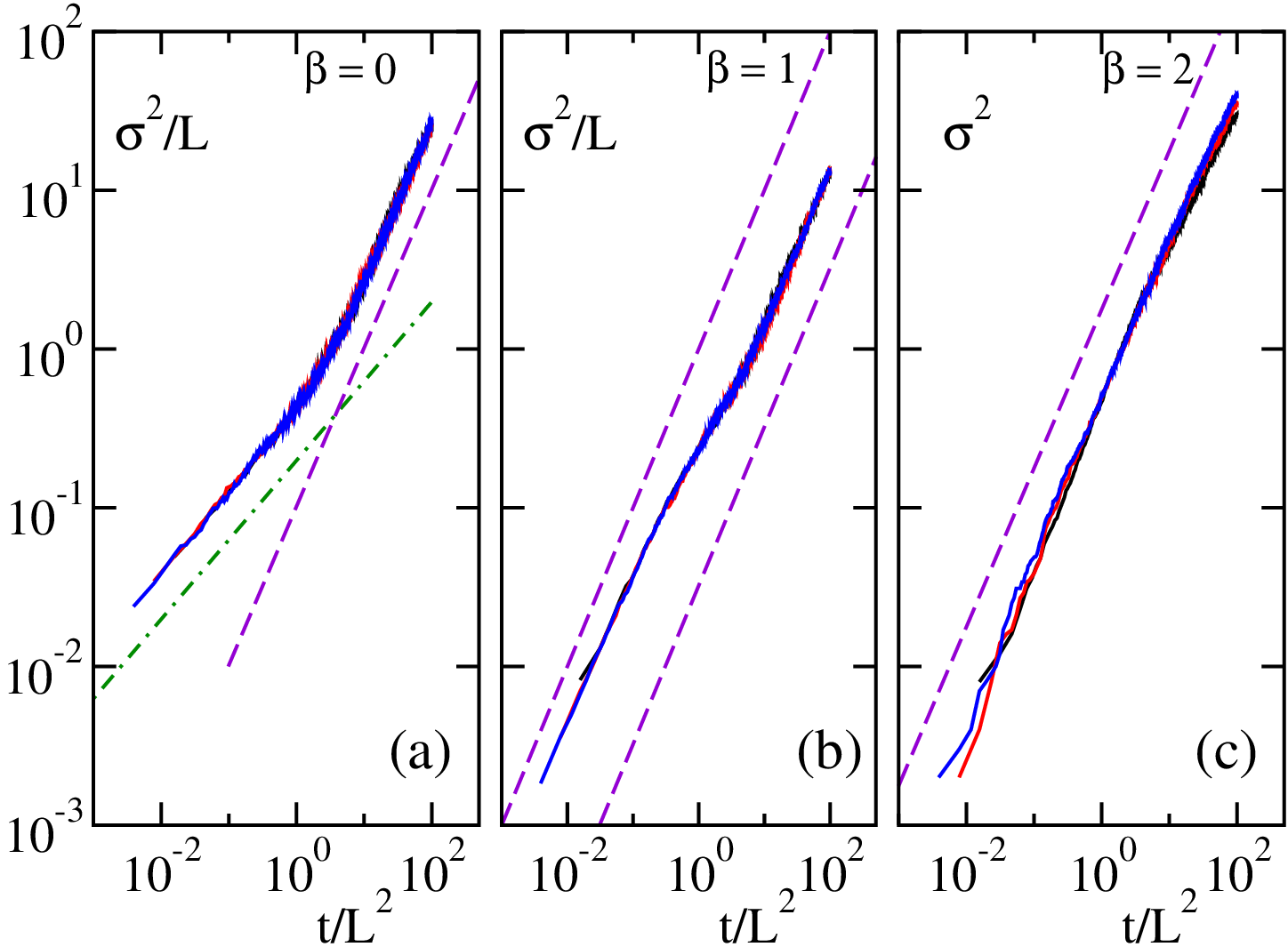}
\end{center}
\caption{
Evolution of the amplitude fluctuations $\sigma^2$ of two breathers sitting on 
an infinite temperature background for different coupling exponents $\beta$.
Black, red, and blue full lines (indistinguishable) correspond to $N=64$, 128, and 256, respectively.
Data are obtained by averaging over a set of $1000$ independent trajectories.
The initial condition consists of two breathers of equal amplitude $h(0)=N=L$ at the boundaries of the chain. 
Dotted-dashed (green) and dashed (purple) lines correspond to a square root and linear increase of $\sigma^2$,
respectively. 
}  
\label{f:scal_beta}
\end{figure}

\begin{figure}[htbp]
\begin{center}
\includegraphics*[width=0.8\textwidth]{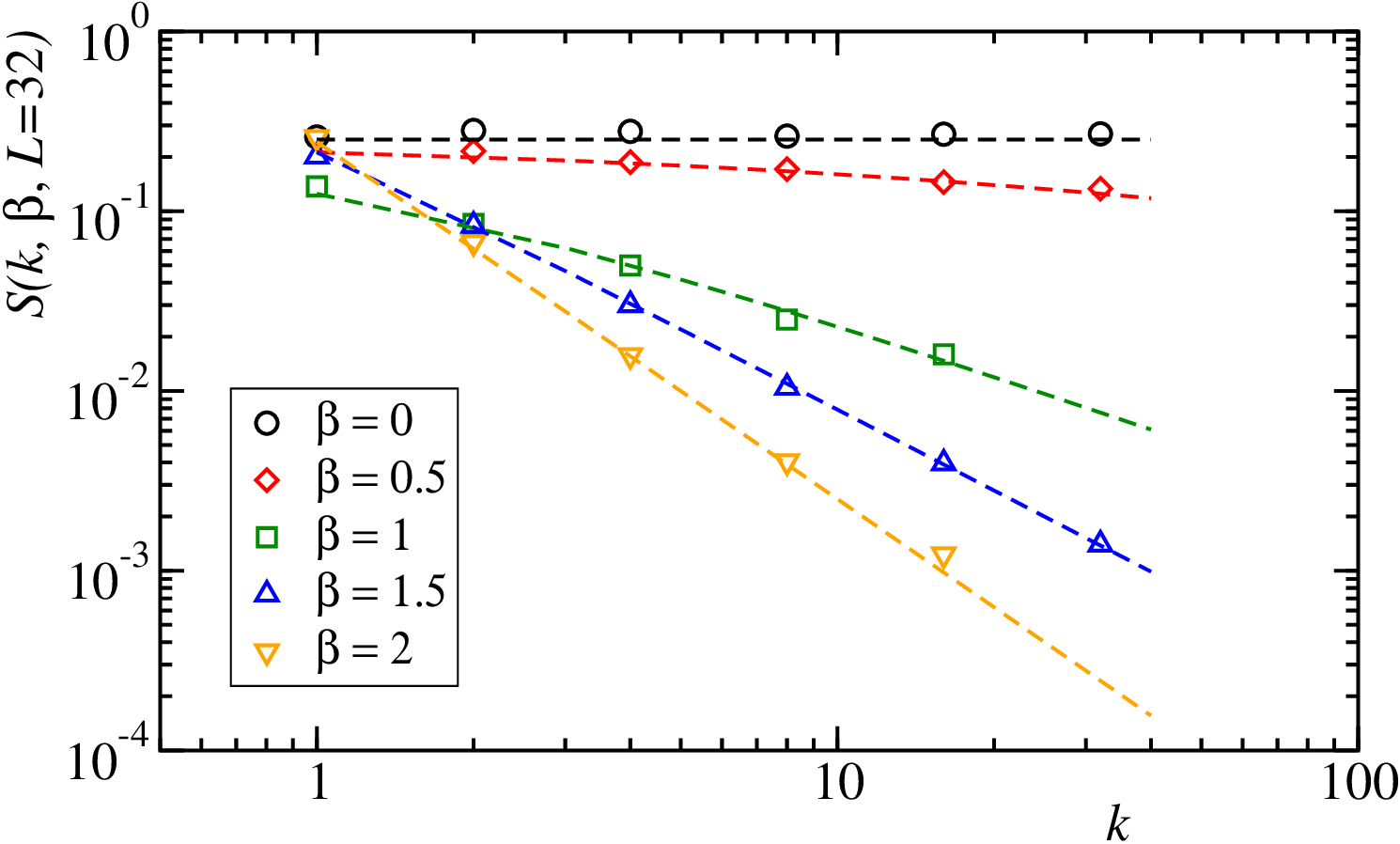}
\end{center}
\caption{
The diffusion prefactor $S(k,\beta,L=32)$ (open symbols) computed from a fit of the asymptotic growth of $\sigma^2(t)$
for different values of $\beta$ and $k$ in a chain of $N=32$ sites.
Dashed lines are obtained from the analytical prediction in Eq.~(\ref{eq:pref}).
Simulations are performed by superposing two breathers with amplitude $h(0)=kL$ at the two
boundaries of  a PEP chain at infinite temperature and with  fixed boundary conditions.
The breather interaction probability is $\alpha=h(0)^{-\beta}$.
For each choice of  $k$ and $\beta$ ,  the mean square height displacement $\sigma^2$
is computed by averaging over 1000 independent trajectories.
The coefficient $S(k,\beta,L=32)$ is therefore extracted by performing a linear fit of $\sigma^2$
in the asymptotic region where $\sigma^2 \sim t$. 
}
\label{f:coeff_diff}
\end{figure}

In Fig.~\ref{f:scal_beta} we show the growth of $\sigma^2(t)$ as obtained by directly simulating the evolution 
of a PEP model where two breathers are initially superposed to a background  in equilibrium at infinite
temperature (i.e., with   a density $1/2$).
Horizontal and vertical axes are rescaled so as to collapse  data corresponding to different system sizes on a single curve.
In all cases, the growth of $\sigma^2$ is asymptotically linear: $\sigma^2(t)\sim t/L$ for $\beta=0$ 
and $\beta=1$, while $\sigma^2(t) \sim t/L^2$ for $\beta=2$, in agreement  with the prediction in Eq.~(\ref{eq:diff_const}).
A detailed check of the prefactor $S(k,\beta,L)$ extracted from numerical simulations, is presented in
Fig.~\ref{f:coeff_diff}, where it is compared with the analytical formula, Eq.~(\ref{eq:pref}).
An excellent agreement is found.

It is interesting to notice that the asymptotic linear growth of $\sigma^2$, which confirms the eventual diffusive
behavior of the breather amplitude, may be preceeded by a sub-diffusive behavior, where $\sigma \simeq t^{1/4}$
(see the green dotted curve in Fig.~\ref{f:scal_beta}). This behavior can be understood by invoking a
somehow unexpected relationship with surface roughening phenomena~\cite{KRB}.
Consider a finite PEP extending from site 0, where it is in
contact with a breather, to site $L$ where fixed boundary conditions are imposed.
Let $c_i(t)$ be a variable denoting whether a particle is present or not on site $i$ at time $t$ and introduce
\[
C_j(t) = \sum_{i\ge j}^L c_i(t)  \; .
\]
The variable $C_j(t)$ counts the number of particles present in the system on the right of the site $j$.
It can be written as $C_j(t) = (L-j)/2 + s_j(t)$, where $s_j(t)$ can be interpreted as a rough interface
of vanishing average height.
Therefore the fluctuations of $s_0(t)$ (and thereby of $C_0(t)$), represent the fluctuations of the breather height 
as well as the fluctuations of the rough surface. In the contexts analysed in this paper the bulk dynamics is
fully linear, so that it is appropriate to invoke an analogy with the Edwards-Wilkinson model, whose
fluctuations are precisely characterized by the exponent 1/4~\cite{EW1982} seen in Fig.~\ref{f:scal_beta}
for $\beta=0$. 

This ``early time" behavior appears because in between the emission of a particle from a breather 
and the next emission, the system may not have the time to reach local equilibrium when $\beta$ is small.
$\beta=1$ is the limiting value for the observation of the initial slow growth. In this critical case, the anomaly
reduces to just a different multiplicative coefficient in front of the linear term, see Fig.~\ref{f:scal_beta}(b).

We conclude this Section with the remark that above derivation of $\tau(L)$ applies to any $\alpha(L)$
dependence so long as it represents a coupling which weakens upon increasing the breather height.
This includes an exponential dependence, in which case the time required for the death of a breather 
grows exponentially with its size.

\section{The MMC model}

Here, we discuss the coarsening process in the MMC setup. As briefly anticipated in the introduction,
the evolution rule is a typical microcanonical Monte Carlo move restricted to
neighbouring particles, so as to maintain the locality of the interactions of the DNLS equation. 
In practice, a triplet of neighbouring sites, $(n-1,n,n+1)$, is
randomly chosen and the variables $(a_{n-1},a_n,a_{n+1})$ updated so as to conserve mass and energy.

The positivity of $a_n$ implies that when a high-mass (breather) site is involved 
(see Fig.~\ref{f:MMC_PEP}(a)), the accessible phase space reduces, to the point that,
if a finite mass were concentrated in a single site, the two neighbouring ones 
being perfectly empty, no redistribution would be possible at all. 
%As a consequence of such a restriction a breather phase emerges, characterized by coarsening
%the emergence of coarsening phenomena when $u>2a^2$~\cite{Rumpf2008,IPP14}.
In the absence of the condition for the mass to be positive, the rule would be equivalent to
a stochastic scheme which preserves kinetic energy and linear momentum, of the type used
to ``ergodize'' chains of oscillators~\cite{BBO06,Lepri2009,Lepri2010,Delfini10}, where no such exotic
phenomena are observed.

The same analysis carried out in the previous sections for the PEP, could be repeated for the MMC,
by introducing a suitable height-dependent coupling. We have verified that this leads to
the same scenario and, therefore, we do not see a compelling reason to show the corresponding results.
We rather propose some considerations which make the weakening assumption less ad hoc than
introducing a priori a dependence of the probability $\alpha$ on the breather height.

We start by quickly reminding that in the MMC model~\cite{IPP14} we choose a random triplet of neighboring sites
and update their amplitudes under the constraint of constant mass, $a_{n-1} +a_n +a_{n+1}= \bar a$,
and constant energy, $a_{n-1}^2 +a_n^2 +a_{n+1}^2 = \bar u$. These two conditions correspond to the
intersection of a sphere with a plane in the three dimensional phase space, $a_{n-1},a_n,a_{n+1}$.
Since $a_n$ represent a mass it must be positive. This implies that the intersection is a full circle if
the initial amplitudes are comparable, while it is made up of three, disconnected arcs if one of the sites has
an initial amplitude significantly larger than the others (see Fig.~\ref{f:MMC_PEP}(a)). 
In practice, this occurs when one of the three sites is occupied by a breather. 

A first consequence of the results of the previous section is that different microscopic rules for the
evolution of the system may produce the same coarsening exponent $1/\zeta$.
More specifically, we remind that in Ref.~\cite{IPP14} the MMC rule was implemented by restricting the random
selection to the fully positive triplets, i.e. choosing points which lie inside the allowed
arcs (this could be the entire circle). Here we consider a possible variation of the above dynamics that
amounts to always selecting a point along the full circle and accepting the move only if the positivity condition 
is satisfied for the three amplitudes. Compared to the former recipe, which corresponds to  $\alpha =1$, i.e. $\beta =0$,
the latter algorithm avoids the identification of the arc extrema as functions of the mass and energy 
of the triplet. On the other hand, the dynamics is slowed down whenever the algorithm generates a triplet
with at least one negative amplitude. Therefore one expects that this change or rule affects the probability $\alpha$ 
to effectively perturb a breather.

We now show that the slowing down corresponds to $\alpha \sim L^{-1/2}$, i.e. $\beta=1/2$.
Consider a triplet that contains one breather site with amplitude
$b$ and two background sites with amplitude $xb$ and $yb$, with $x,y\ll 1$ and let $\epsilon$ denote
the angular length of each arc\footnote{By symmetry reasons the three arcs have
the same length $\epsilon$. When $\epsilon$ equals $2\pi/3$, they merge into the entire circle.}.
Then, the probability $P_a$ that a randomly chosen move is acceptable is given by the ratio 
$3\epsilon/2\pi$, where the denominator is the amplitude of the set of moves that conserve mass and energy
and the numerator accounts for the amplitude of the physical interval (i.e. the set of moves that satisfy 
also positivity). Since to leading order $\epsilon=\sqrt{3}(x+y)$~\cite{IPP14}, 
$\epsilon$ scales as $1/b$ when the background amplitude is kept fixed and then also $P_a\sim 1/b$.
Moreover, considering that the quantity diffusing during the MMC dynamics is the energy
$u=b^2$~\cite{IPP14}, $P_g\sim u^{-1/2}$, so that $\alpha \sim L^{-1/2}$.
According to the discussion of previous Section, 
the difference between the two $\beta$ exponents, $\beta=0$ and $\beta=1/2$, is not sufficient
to induce a different coarsening law, which is again characterized by the exponent $\zeta=1/3$ (data not shown).

Finally, we discuss a modification of the model which naturally leads to an exponentially slow dynamics,
without the need to introduce explicitly an exponential decrease of $\alpha$ with $L$.
In practice, we introduce a threshold $\Gamma$ for the minimal arc angle to be considered.
In other words, whenever the arc angle $\epsilon$ is smaller than $\Gamma$, no action is taken: the triplet configuration is left
unchanged.

\begin{figure}[htbp]
\begin{center}
\includegraphics[width=0.8\textwidth,clip]{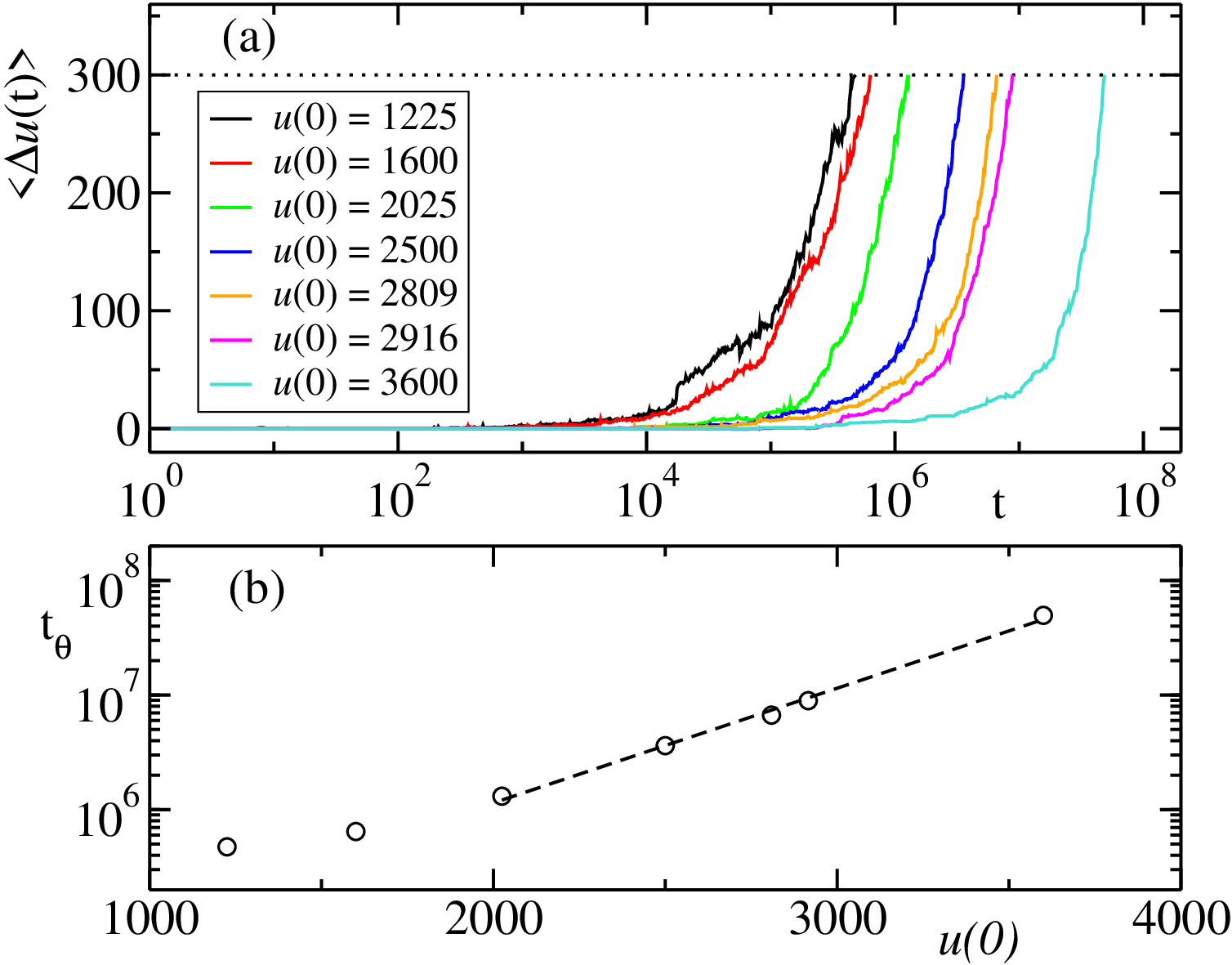}
\end{center}
\caption{(a): Average energy reduction $\langle \Delta u(t)\rangle\equiv \langle u(0)-  u(t) \rangle$ 
of one breather in a modified MMC model with interaction threshold $\Gamma=\pi/8$
for different initial energies $u(0)$.
At $t=0$ the breather is superposed to an MMC chain in equilibrium at temperature $T=10$ with mass density $a=1$ 
and $L=2048$. The initial  equilibrium distribution of the background is sampled by applying a Metropolis
thermostat at $T=10$ to each site of the chain for a transient $t_{th}=5\cdot 10^5$. 
$\langle \Delta u(t)\rangle$ is computed by averaging over $40$ independent trajectories
(up to the threshold $\theta=300$).
(b): Relaxation time (open circles) $t_\theta$ to reach the threshold as a function of the initial breather 
energy $u(0)$. The dashed line refers to an exponential fit $t_\theta \sim \exp(u(0)/E)$ with $E=430$.
}
\label{f:exp_rel}
\end{figure}

Once again, let us analyse the implications of the algorithm when the chosen triplet contains one breather 
and two background sites with amplitude $b,xb,yb$ respectively.
%with amplitude $b$ and two background sites with amplitude $xb$ and $yb$, with $x,y\ll 1$. To leading order, 
%$\lambda=\sqrt{3}(x+y)$~\cite{IPP14}.
So long as both $x$ and $y$ are small 
the dynamics is blocked, since $\epsilon=\sqrt{3}(x+y)<\Gamma$.
One may naively conclude that sufficiently high breathers are completely decoupled from the background.
This is not true, because the background fluctuations can eventually lead to sufficiently
large $x$ or $y$ values, so that the probability of such move is related to the probability of generating
sufficiently large amplitudes in the neighbouring sites. 
More precisely, let us consider the
grand canonical equilibrium distribution of the background amplitudes $a_n$ reads 
\be
P(a_n)=\frac{1}{Z}\exp[-\beta(a_n^2-\mu a_n)], 
\ee
where $\beta=1/T$ and $\mu$ are the inverse (positive) temperature and the chemical potential 
of the background, while $Z=\int_0^\infty\exp[-\beta(a_n^2-\mu a_n)]\,da_n$ is the partition function. 
The probability to have a mass fluctuation comparable to the 
breather height $b$ (the square root of the  energy) 
is exponentially small in $b$ and depends on the values of $\beta$ and $\mu$.
In fact, a direct simulation of the modified MMC model with $\Gamma=\pi/8$ (see Fig.~\ref{f:exp_rel}(a)), 
shows that breathers well above the interaction threshold evolve according to an effective coupling exponentially 
small in the breather energy.

With reference to the parameters of Fig.~\ref{f:exp_rel}, $\beta=0.1$ and $\mu=-6.46$
\footnote{The chemical potential $\mu$ has been determined by solving the equation 
$a=\int_0^\infty P(x)x\,dx$ with $\beta=0.1$ and $a=1$.
}. In the range of
chosen breather energies ($u(0)>1000$), the term in $\mu$ in $P(a_n)$ can be neglected, so that the probability of
the fluctuation is controlled by the local energy $a_n^2$. This is confirmed in Fig.~\ref{f:exp_rel}(b), where 
the relaxation time $t_\theta$ necessary to release an amount of energy $\theta$ from the breather to the
background is shown to depend exponentially on the initial breather energy $u(0)$.
Altogether, the above analysis shows that an exponentially small coupling between breathers and
background may arise as a consequence of background fluctuations when energy diffusion is
blocked by additional constraints. In this regime, the analysis of the previous section
predicts a logarithmic coarsening of the breathers.  

\section{Conclusions}

The deterministic DNLS equation and the stochastic MMC/PEP models
have the common feature that their phase diagram includes a high-energy region
where localized excitations can appear (DNLS and MMC) or can be preserved (PEP), at least.
Breather dynamics is frozen in DNLS, while it is normally slowed down, but not frozen,
in MMC/PEP. The slow dynamics in stochastic models corresponds to a coarsening
process where breathers exchange mass until a breather loses so much mass that
it is absorbed by the background, leading to a reduction of breather density.

This coarsening process follows a power-law and it is more or less slow, depending
on the strength of the breather-background interaction.
Our results support the idea that the extremely slow (almost frozen) dynamics observed in DNLS is 
due to the weakening of the interaction with breathers of increasing height.
However, a power law weakening of the breather-background coupling maintains a power-law
coarsening. For this reason, the results of section 4 are particularly instructive.
There, we have studied a variant of the  MMC model, where the breather-background interaction is
regulated by a threshold rather than by a direct coupling: in fact, we impose there is
a minimum trasfer of mass. Upon increasing the breather height,
this condition can be satisfied only if a neighbouring site is substantially higher than usual,
an event that is exponentially rare and highly intermittent.

This mechanism, postulated a priori in the MMC model, might be active in the DNLS as a result
of the combination of the usually very weak interaction of high-amplitude breathers 
(due to their fast rotation) with rare, relatively-strong, energy transfers due to
occasional resonant interactions with neighbouring sites.
A detailed investigation of this conjecture is in progress.

\appendix

\section{Diffusion with a single semi-reflecting boundary}
\label{app.Robin}

We want to solve the diffusion equation,
\be
\frac{\partial p}{\partial t} = D \frac{\partial^2 p}{\partial x^2} ,
\label{ea.diff}
\ee
in the semi-infinite line $x\ge 0$, with a semi-reflecting boundary in $x=0$,
\be
\left. \frac{\p p}{\p x} \right|_{x=0} = r p(x=0,t) ,
\ee
and with initial condition $p(x,t=0) = \delta(x-x_0)$. We remind that within the PEP model,
we have $r=2\alpha$, see Eq.~(\ref{eq:robin}).

We define $q(x,t) = p'(x,t) - rp(x,t)$, where the prime indicates the derivative with respect to $x$,
which still satisfies the diffusion equation (\ref{ea.diff}), but with the easier boundary condition,
$q(0,t)=0$. The price to be paid is in the initial condition, $q(x,0) =\delta'(x-x_0) -r\delta(x-x_0)$.
The details of the calculation can be found in Ref.~\cite{Weiss} and here we limit to write the
solution
\bea
\nonumber
q(x,t) &=& -\frac{1}{\sqrt{4\pi (Dt)^3}}
\left[ 
(x+x_0) e^{ -\frac{(x+x_0)^2}{4Dt}} +(x-x_0) e^{-\frac{(x-x_0)^2}{4Dt}}
\right] \\
&& -\frac{r}{\sqrt{4\pi Dt}} 
\left[
e^{-\frac{(x-x_0)^2}{4Dt}} -  e^{-\frac{(x+x_0)^2}{4Dt}}
\right]
\label{ea.q}
\eea
and 
\be
p(x,t) = -\int_x^\infty q(s,t) e^{-r(s-x)} ds .
\ee

What we need is the probability $F(t)$ that the particle is absorbed in $x=0$ during the time
interval $(t,t+dt)$, which is given by
\be
F(t) = Dp'(0,t) = Dr p(0,t)  
= -Dr \int_0^\infty q(x,t) e^{-rx} dx .
\ee
Therefore, using (\ref{ea.q}) we obtain
\bea
\nonumber
\hskip -2.cm F(t) &=& Dr\int_0^\infty dx e^{-rx} \left\{ \frac{1}{\sqrt{4\pi (Dt)^3}}
\left[ 
(x+x_0) e^{-\frac{(x+x_0)^2}{4Dt}} +(x-x_0)e^{-\frac{(x-x_0)^2}{4Dt}}
\right] \right.\\
\hskip -2.cm && \left. +\frac{r}{\sqrt{4\pi Dt}} 
\left[
e^{-\frac{(x-x_0)^2}{4Dt}} - e^{-\frac{(x+x_0)^2}{4Dt}}
\right] \right\} .
\eea

We now formally evaluate the previous integral, which may be written in terms of the
error function, $\erf(x) = \frac{2}{\sqrt{\pi}}\int_0^x e^{-s^2} ds$.
Then, we evaluate the limiting behaviors of $F(t)$, at short and long times.
We need to calculate the following integrals,
\bea
I_1(x_0) &=& \int_0^\infty dx e^{-rx} e^{-\frac{(x+x_0)^2}{4Dt}} \\
&=& e^{rx_0} \sqrt{\pi Dt} \; e^{r^2Dt} \left[ 1-\erf\left( r\sqrt{Dt} +\frac{x_0}{2\sqrt{Dt}}\right)\right] \\
&=& e^{rx_0} \sqrt{\pi Dt} \; J_1(x_0,r,t)
\eea
where
\be
J_1(x_0,r,t) \equiv e^{r^2Dt} \left[ 1-\erf\left( r\sqrt{Dt} +\frac{x_0}{2\sqrt{Dt}}\right)\right]
\ee
and
\bea
I_2(x_0) &=& \int_0^\infty dx e^{-rx} (x+x_0) e^{-\frac{(x+x_0)^2}{4Dt}} \\
&=& -e^{rx_0} \sqrt{\pi Dt} \frac{\partial}{\partial r} J_1(x_0,r,t) ,
\eea
so that
\bea
F(t) &=&  -\frac{r}{2t} \left[ e^{rx_0} \partial_r J_1(x_0,r,t) + e^{-rx_0} \partial_r J_1(-x_0,r,t)\right] \\
&& +\frac{Dr^2}{2} \left[ e^{rx_0} J_1(x_0,r,t) - e^{-rx_0} J_1(-x_0,r,t)\right] .
\eea

$\bullet$ Limit $x_0^2 \ll Dt \ll r^{-2}$

Using the limit $\erf(x) \simeq \frac{2}{\sqrt{\pi}} x$ for $x\ll 1$, we can write
\bea
J_1(x_0,r,t) &\simeq& 1 -\frac{2}{\sqrt{\pi}} \left( r\sqrt{Dt} + \frac{x_0}{2\sqrt{Dt}} \right) \\
\partial_r J_1(x_0,r,t) &\simeq& - \frac{2}{\sqrt{\pi}} \sqrt{Dt} 
\eea
and
\be
F(t) \simeq \frac{2}{\sqrt{\pi}} \sqrt{D} \frac{r}{\sqrt{t}} .
\label{Ft1}
\ee

$\bullet$ Limit $t\gg r^{-2}$

Using the limit $\erf(x) \simeq 1 -\frac{e^{-x^2}}{\sqrt{\pi}x}$ for $x\gg 1$, we can write

\bea
J_1(x_0,r,t) &\simeq&  \frac{1}{r\sqrt{\pi D t}}\\
\partial_r J_1(x_0,r,t) &\simeq& -\frac{1}{r^2 \sqrt{\pi D t}}
\eea
and
\be
F(t) \simeq \frac{1}{r\sqrt{\pi D}} t^{-3/2} .
\label{Ft2}
\ee

Using $r=2\alpha$, from (\ref{Ft1}) and (\ref{Ft2}), we find Eqs.~(\ref{eq:F(t)}).

\section{Relaxation of a breather on an empty background}
\label{app.relax}

%\be
%\Delta h(t) = \alpha (1-p_1) \int_0^t dt'\int_{t'}^{\infty} dt'' F(t'').
%\ee
$\bullet$ Limit $x_0^2\ll t\ll\alpha^{-2}$

Recalling that 
\be
\int_{0}^\infty dt F(t) =1\quad,
\ee
we rewrite Eq.~(\ref{eq:h(t)}) as
\be
\langle \Delta h(t) \rangle = \alpha \int_0^t dt' (1-p_1(t'))\left[1-\int_{0}^{t'} dt'' F(t'')\right].
\ee
Now, since at short times
\be
\int_0^{t'}  dt'' F(t'') \ll D^{1/2} \simeq 1 
\ee
and $p_1\ll 1$,
we obtain to the leading order
\be
\langle \Delta h(t) \rangle= \alpha t.
\ee

$\bullet$ Limit $t\gg\alpha^{-2}$

Let us introduce an intermediate timescale $t_0$ such that $\alpha^{-2}\ll t_0\leq t$.
Therefore, Eq.~(\ref{eq:h(t)}) is rewritten as
\be
\hskip -2.5cm \langle \Delta h(t) \rangle= \alpha \!\left[ \int_0^{t_0}\! \!dt' (1-p_1(t'))\!\int_{0}^{t'}\!\! dt'' F(t'') +
\int_{t_0}^{t}\!\! dt' (1-p_1(t'))\! \int_{0}^{t'}\!\! dt'' F(t'')\right].
\ee
In the regime $t\gg \alpha^{-2}$, $F(t)$ can be approximated as in Eq.~(\ref{eq:F(t)}) and $p_1(t)\simeq 1/2$.
Consequently, one can compute explicitly the second addend in the square brackets, that gives
\be
\hskip -2.cm \int_{t_0}^{t} dt' (1-p_1(t'))\int_{0}^{t'} dt'' F(t'') = \frac{1}{2} 
\int_{t_0}^{t} dt'\int_{0}^{t'} dt'' \frac{1}{2\alpha\sqrt{\pi Dt''^3}}
=\frac{1}{\alpha}\sqrt{\frac{t}{\pi D}}.
\ee
Therefore, for $t \geq t_0$
\be
\langle \Delta h(t) \rangle= K(t_0) + \sqrt{\frac{t}{\pi D}},
\ee
where $K(t_0)$ is a constant. Finally, for $D=1/2$, we obtain Eq.~(\ref{PEP_large_t}).

\section{Random walk with two semi-reflecting boundaries}
\label{app.rw}

We have two breathers in $i=0,L$ and a random walk in between, moving according to the following rules:
if the particle is in $i\ne 1,L-1$, it hops to $i\pm 1$ with probability $1/2$; if $i=1$ ($i=L-1$), 
it hops to $i=2$ ($i=L-2$) with probability $q$ and to
$i=0$ ($i=L$), therefore being absorbed, with probability $1-q$.
We define the probability $P(L)$ that a particle released in $i=1$ attaches to the breather in $i=L$.

It is also useful to define a model where the left boundary condition is the same as before, but
the right boundary condition is symmetric, i.e.
the particle hops from $i=L-1$ to $i=L-2,L$ with probability $1/2$. 
For this model we define the probability $p(L)$ that a particle released in $i=1$ reaches the site $i=L$
before reaching the site $i=0$.  With these notations, we can write
\be
P(L) = p\left( \frac{L}{2}\right) \frac{1}{2} ,
\ee
because a particle arrived in $L/2$ has the same probability to attach to the left or to the right breather.
We now want to determine $p(L)$, summing up all the possible trajectories to go from $i=1$ to $i=L$, without
being absorbed in $i=0$, according to the number $n$ of passages in $i=1$ (with $n\ge 1$).
If a trajectory is characterized by a given $n$, it means that the particle has hopped from 1 to 2 (which
happens with probability $q$) and $(n-1)$ times it has come back to 1 before attaining $L$ (which occurs
with probability $(1-a)$, with $a=1/L$), and one time has attained $L$ before coming back to 1 (which occurs
with probability $a$). Summing up all terms, we have
\bea
p(L) &=& \sum_{n\ge 1} q^n (1-a)^{n-1} a \\
&=& \frac{a}{1-a} \sum_n q^n (1-a)^n \\
&=& \frac{aq}{(1-q) + qa} ,
\eea
with $a=1/L$. So, we get
\bea
P(L) &=& p\left( \frac{L}{2}\right) \frac{1}{2} \\
&=& \frac{q}{2q + L(1-q)} \\
&=& \frac{1}{2 +L(q^{-1} -1)} .
\eea

Let us now define $q$ in the most general way.
If a particle is in $i=1$, it has 
a probability $1/4$ to move to the right and a probability $\alpha/2$ to move to the left,%
\footnote{Each probability is the product of $1/2$ (the probability to choose right or left)
with the probability to actually make the move.}
so 
\be
q = \frac{\frac{1}{4}}{\frac{1}{4} +\frac{\alpha}{2}} = \frac{1}{1+2\alpha} ,
\ee
and
\be P(L) = 
\frac{1}{2}\frac{1}{1+L\alpha} .
\ee

\section*{References}

\bibliographystyle{iopart-num}

\providecommand{\newblock}{}

\end{document}